\documentclass[sigconf,balance=true,hidelinks,table]{acmart}

\usepackage{hyperref}
\usepackage{amsmath,amsfonts}
\usepackage{algorithmic}
\usepackage[ruled,linesnumbered,noend]{algorithm2e}
\usepackage{xcolor}
\usepackage{graphicx}
\usepackage{textcomp}
\usepackage{xcolor}
\usepackage{nccmath}
\usepackage{environ}
\usepackage{tcolorbox}
\usepackage{soul}
\usepackage{booktabs}
\usepackage{multirow}
\usepackage{array}
\usepackage{float}
\usepackage{makecell}
\usepackage{flushend}
\usepackage{enumitem}
\usepackage{hhline}
\usepackage{comment}
\usepackage{epstopdf} 
\usepackage{caption}
\usepackage{subcaption}

\usepackage{xcolor}
\usepackage{listings}

\NewEnviron{myequation}{%
\begin{equation}
\scalebox{1}{$\BODY$}
\end{equation}
}

\newcommand{\fig}{Fig.}
\newcommand{\tab}{Table}

\newcolumntype{P}[1]{>{\centering\arraybackslash}p{#1}}

\newcommand{\code}[1]{\textnormal{\texttt{#1}}}

\acmConference[MSR 2022]{MSR '22: Proceedings of the 19th International Conference on Mining Software Repositories}{May 23–24, 2022}{Pittsburgh, PA, USA}

\begin{document}

\title{On the Use of Fine-grained Vulnerable Code Statements for Software Vulnerability Assessment Models}

\author{Triet Huynh Minh Le}
\affiliation{\institution{CREST - The Centre for Research on Engineering Software Technologies, The University of Adelaide}
\city{Adelaide}
\country{Australia}}
\email{triet.h.le@adelaide.edu.au}

\author{M. Ali Babar}
\affiliation{\institution{CREST - The Centre for Research on Engineering Software Technologies, The University of Adelaide}
\city{Adelaide}
\country{Australia}}
\affiliation{\institution{Cyber Security Cooperative Research Centre, Australia}
\city{}
\country{}}
\email{ali.babar@adelaide.edu.au}

\begin{abstract}
Many studies have developed Machine Learning (ML) approaches to detect Software Vulnerabilities (SVs) in functions and fine-grained code statements that cause such SVs.
However, there is little work on leveraging such detection outputs for data-driven SV assessment to give information about exploitability, impact, and severity of SVs.
The information is important to understand SVs and prioritize their fixing.
Using large-scale data from 1,782 functions of 429 SVs in 200 real-world projects, we investigate ML models for automating function-level SV assessment tasks, i.e., predicting seven Common Vulnerability Scoring System (CVSS) metrics.
We particularly study the value and use of vulnerable statements as inputs for developing the assessment models because SVs in functions are originated in these statements.
We show that vulnerable statements are 5.8 times smaller in size, yet exhibit 7.5-114.5\% stronger assessment performance (Matthews Correlation Coefficient (MCC)) than non-vulnerable statements.
Incorporating context of vulnerable statements further increases the performance by up to 8.9\% (0.64 MCC and 0.75 F1-Score).
Overall, we provide the initial yet promising ML-based baselines for function-level SV assessment, paving the way for further research in this direction.

\end{abstract}



\maketitle

\section{Introduction}

Software Vulnerabilities (SVs) can make software systems susceptible to catastrophic cyber-attacks. Thus, addressing such SVs is a critical task for developers. However, fixing all SVs at the same time is not always practical due to limited resources and time~\cite{khan2018review}. A common practice in this situation is to prioritize fixing SVs posing imminent and serious threats to a system of interest. Such prioritization usually requires inputs from SV assessment~\cite{le2021survey}, and Common Vulnerability Scoring System (CVSS)~\cite{cvss_website} is one of the most popular SV assessment frameworks. CVSS characterizes SVs using various metrics including their exploitability, impact, and severity. The metrics can then be used to select critical SVs to fix early. For example, a severe SV that is easily exploitable and has critical impacts on a system would likely have a high fixing priority.

Previous studies (e.g.,~\mbox{\cite{yamamoto2015text,han2017learning,spanos2018multi,le2019automated,le2021survey}}) have mostly used SV reports to develop data-driven models for assigning the CVSS metrics to SVs. Among sources of SV reports, National Vulnerability Database (NVD)~\cite{nvd_website} has been most commonly used for building SV assessment models~\cite{le2021survey}. The popularity of NVD is mainly because it has SV-specific information (e.g., CVSS metrics) and less noise in SV descriptions than other Issue Tracking Systems (ITSs) like Bugzilla~\cite{bugzilla}. The discrepancy is because NVD reports are vetted by security experts, while ITS reports may be contributed by users/developers with limited security knowledge~\cite{croft2021investigation}.
However, NVD reports are mostly released long after SVs have been fixed.
Our analysis revealed that less than 3\% of the SV reports with the CVSS metrics on NVD had been published before SVs were fixed; on average, these reports appeared 146 days later than the fixes. Note that our findings accord with the previous studies~\cite{li2017large,piantadosi2019fixing}. This delay renders the CVSS metrics required for SV assessment unavailable at fixing time, limiting the adoption of report-level SV assessment for understanding SVs and prioritizing their fixes.

Instead of using SV reports, an alternative and more straight-forward way is to directly take (vulnerable) code as input to enable SV assessment prior to fixing.
Once a code function is confirmed vulnerable, SV assessment models can assign it the CVSS metrics before the vulnerable code gets fixed, even when its report is not (yet) available.
Note that it is non-trivial to use static application security testing tools to automatically create bug/SV reports from vulnerable functions for current SV assessment techniques as these tools often have too many false positives~\mbox{\cite{johnson2013don,aloraini2019empirical}}.
To develop function-level assessment models, it is important to obtain input information about SVs in functions detected by manual debugging or automatic means like data-driven approaches (e.g.,~\mbox{\cite{zhou2019devign,zheng2020impact,lin2020deep}}).
Notably, recent studies (e.g.,~\mbox{\cite{li2021vulnerability,nguyen2021information}}) have shown that an SV in a function usually stems from a very small number of code statements/lines, namely \textit{vulnerable statements}.
Intuitively, these vulnerable statements potentially provide highly relevant information (e.g., causes) for SV assessment models. Nevertheless, a large number of other (non-vulnerable) lines in functions, though do not directly contribute to SVs, can still be useful for SV assessment, e.g., indicating the impacts of an SV on nearby code.
It still remains largely unknown about function-level SV assessment models as well as the extent to which vulnerable and non-vulnerable statements are useful as inputs for these models.

We conduct a large-scale study to fill this research gap. We investigate the usefulness of integrating fine-grained vulnerable statements and different types of code context (relevant/surrounding code) into learning-based SV assessment models. The assessment models employ various feature extraction methods and Machine Learning (ML) classifiers to predict the seven CVSS metrics (Access Vector, Access Complexity, Authentication, Confidentiality, Integrity, Availability, and Severity) for SVs in code functions.

Using 1,782 functions from 429 SVs of 200 real-world projects, we evaluate the use of vulnerable statements and other lines in functions for developing SV assessment models. Despite being up to 5.8 times smaller in size (lines of code), vulnerable statements are more effective for function-level SV assessment, i.e., 7.4-114.5\% higher Matthews Correlation Coefficient (MCC) and 5.5-43.6\% stronger F1-Score, than non-vulnerable lines. Moreover, vulnerable statements with context perform better than vulnerable lines alone.
Particularly, using vulnerable and all the other lines in each function achieves the best performance of 0.64 MCC (8.9\% better) and 0.75 F1-Score (8.5\% better) compared to using only vulnerable statements.
We obtain such improvements when combining vulnerable statements and context as a single input based on their code order, as well as when treating them as two separate inputs.
Having two inputs explicitly provides models with the location of vulnerable statements and context for the assessment tasks, while single input does not.
Surprisingly, we do not obtain any significant improvement of the double-input models over the single-input counterparts.
These results show that function-level SV assessment models can still be effective even without knowing exactly which statements are vulnerable.
Overall, our findings can inform the practice of building function-level SV assessment models.

Our key \textbf{contributions} are summarized as follows:

\begin{itemize}[noitemsep,topsep=0pt]
    \item To the best of our knowledge, we are the first to leverage data-driven models for automating function-level SV assessment tasks that enable SV prioritization/planning prior to fixing.
    \item We study the value of using fine-grained vulnerable statements in functions for building SV assessment models.
	\item We empirically show the necessity and potential techniques of incorporating context of vulnerable statements to improve the assessment performance.
    \item We release our datasets and models for future research~\cite{reproduction_package_msr2022}.
\end{itemize}

\noindent \textbf{Paper structure}. Section~\mbox{\ref{sec:background}} gives a background on function-level SV assessment. Section~\mbox{\ref{sec:rqs}} introduces and motivates the three RQs. Section~\mbox{\ref{sec:method}} describes the methods used for answering these RQs. Section~\mbox{\ref{sec:results}} presents our empirical results. Section~\mbox{\ref{sec:discussion}} discusses the findings and threats to validity. Section~\mbox{\ref{sec:related_work}} mentions the related work. Section~\mbox{\ref{sec:conclusions}} concludes the study and suggests future directions.

\section{Background and Motivation}
\label{sec:background}
\subsection{SV Assessment with CVSS}
\label{subsec:cvss_metrics}

Software Vulnerability (SV) assessment is an important step in the SV lifecycle, which determines various characteristics of detected SVs~\cite{smyth2017software}. Such characteristics support developers to understand the nature of SVs, which can inform prioritization and remediation strategies. For example, if an SV can severely damage the confidentiality of a system, e.g., allowing attackers to access/steal sensitive information, this SV should have a high fixing priority. A fixing protocol to ensure confidentiality can then be followed, e.g., checking/enforcing privileges to access the affected component/data.

Common Vulnerability Scoring System (CVSS)~\cite{cvss_website} is one of the most commonly used frameworks by both researchers and practitioners to perform SV assessment. There are two main versions of CVSS, namely versions 2 and 3, in which version 3 only came into effect in 2015. CVSS version 2 is still widely used as many SVs prior to 2015 can yet pose threats to contemporary systems. For instance, the SV with CVE-2004-0113 first found in 2004 was exploited in 2018~\mbox{\cite{old_sv_exploit}}. Hence, we adopt the assessment metrics of CVSS version 2 as the outputs for the SV assessment models in this study.

CVSS version 2 provides metrics to quantify the three main aspects of SVs, namely exploitability, impact, and severity.
We focus on the \textit{base} metrics because the temporal metrics (e.g., exploit availability in the wild) and environmental metrics (e.g., potential impact outside of a system) are unlikely obtainable from project artifacts (e.g., SV code/reports) alone.
Specifically, the base \textit{Exploitability} metrics examine the technique (Access Vector) and complexity to initiate an exploit (Access Complexity) as well as the authentication requirement (Authentication). The base \textit{Impact} metrics of CVSS focus on the system Confidentiality, Integrity, and Availability. The Exploitation and Impact metrics are used to compute the \textit{Severity} of SVs. Severity approximates the criticality of an SV. Nevertheless, relying solely on Severity may be insufficient because an SV with medium severity may still have high impacts as it is considerably complex to be exploited. It is important to assign a high fixing priority to such an SV as an affected system would face tremendous risks in case of a successful cyber-attack. Therefore, in this study, we consider all the base metrics of CVSS version 2, i.e., Access Vector, Access Complexity, Authentication, Confidentiality, Integrity, Availability, and Severity, for developing SV assessment models.

\captionsetup[lstlisting]{labelsep=period, textfont=footnotesize, labelfont=footnotesize, justification=justified}

\newcommand{\lstbg}[3][0pt]{{\fboxsep#1\colorbox{#2}{\strut #3}}}
\lstdefinelanguage{diff}{
  basicstyle=\ttfamily\small,
  morecomment=[f][\lstbg{red!20}]-,
  morecomment=[f][\lstbg{green!20}]+,
}

\definecolor{difftitle}{HTML}{000099}
\definecolor{diffstart}{HTML}{660099}
\definecolor{diffincl}{HTML}{006600}
\definecolor{diffrem}{HTML}{AA3300}

\definecolor{del_color}{HTML}{FFCCCC}
\definecolor{add_color}{HTML}{CCFFCC}

    

\lstdefinestyle{lst}{
    numbers=left, 
    numberstyle=\scriptsize, 
    numbersep = 5pt,
    framexleftmargin = 0in,
    framexrightmargin = 0in,
    xleftmargin = 0.18in,
    xrightmargin = 0.1in,
    basicstyle=\ttfamily\scriptsize, 
    frame=lines,
    showtabs=true,
    showspaces=true,
    showstringspaces=false,
    literate={\ }{{\ }}1,
    escapeinside={<@}{@>}
}

\lstset{belowskip=-0.05in}

    

 




\begin{figure}
\begin{lstlisting}[language=diff,style=lst]
protected String getExecutionPreamble()
{
    if (getWorkingDirectoryAsString() == null)
    {return null;}
    <@\color{blue}{String dir = getWorkingDirectoryAsString();}@>
    <@\color{blue}{StringBuilder sb = new StringBuilder();}@>
    <@\color{blue}{sb.append("cd");}@>
-   sb.append(unifyQuotes(dir));
+   sb.append(quoteOneItem(dir, false));
    <@\color{blue}{sb.append("\&\&");}@>
    <@\color{blue}{return sb.toString();}@>
}
\end{lstlisting}
     
\caption{A vulnerable function extracted from the fixing commit \textit{b38a1b3} of an SV (CVE-2017-1000487) in the \textit{Plexus-utils} project. Notes: Line 8 is vulnerable. Deleted and added lines are highlighted in red and green, respectively. Blue-colored code lines affect or are affected by line 8 directly.}
\label{fig:ex_sv}
\end{figure}


\begin{figure*}[t]
    \centering
    \includegraphics[width=\textwidth,keepaspectratio]{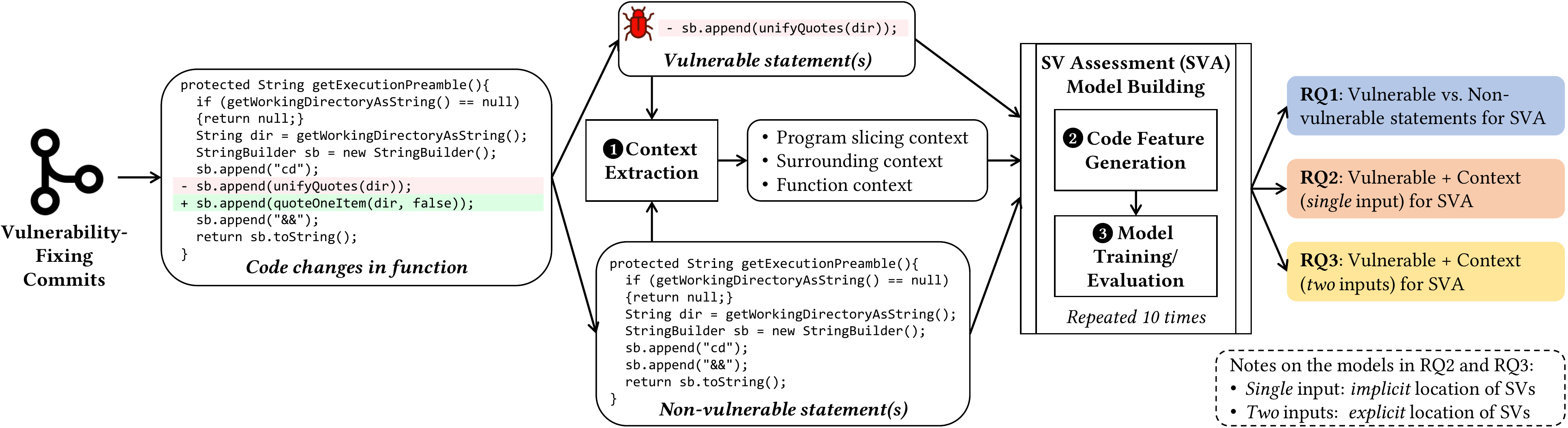}
    \caption{Methodology used to answer the research questions. Note: The vulnerable function is the one described in Fig.~\mbox{\ref{fig:ex_sv}}.}
    \label{fig:workflow}
\end{figure*}

\subsection{SV Assessment in Code Functions}
\label{subsec:vuln_functions}

There have been a growing number of studies to detect vulnerable statements in code functions (e.g.,~\cite{li2021vuldeelocator,li2021vulnerability,nguyen2021information}).
Fine-grained detection assumes that not all statements in a function are vulnerable. We confirmed this assumption; i.e., only 14.7\% of the lines in our curated functions were vulnerable (see section~\ref{subsec:data_collection}). However, it is non-trivial to manually annotate a sufficiently large dataset of vulnerable functions and statements for training SV prediction models. Instead, many studies (e.g.,~\cite{li2021vulnerability,wattanakriengkrai2020predicting,li2020dlfix}) have automatically obtained vulnerable statements from modified lines in Vulnerability Fixing Commits (VFCs) as these lines are presumably removed to fix SVs.
The functions containing such identified statements are considered vulnerable. Note that VFCs are used as they can be relatively easy to retrieve from various sources like National Vulnerability Database (NVD)~\cite{nvd_website}.
An exemplary function and its vulnerable statement are in \fig~\ref{fig:ex_sv}. Line 8 ``\code{sb.append(unifyQuotes(dir));}'' is the vulnerable statement; this line was replaced with a non-vulnerable counterpart ``\code{sb.append(quoteOneItem(dir, false));}'' in the VFC. The replacement was made to properly sanitize the input (\code{dir}), preventing OS command injection.

Despite active research in SV detection, there is little work on utilizing the output of such detection for SV assessment.
Previous studies (e.g.,~\mbox{\cite{yamamoto2015text,spanos2017assessment,spanos2018multi,le2019automated,le2021survey}}) have mostly leveraged SV reports, mainly on NVD, to develop SV assessment models that alleviate the need for manually defining complex rules for assessing ever-increasing SVs.
However, these SV reports usually appear long after SV fixing time.
For example, the SV fix in \fig~\ref{fig:ex_sv} was done 1,533 days before the date it was reported on NVD.
In fact, such a delay, i.e., disclosing SVs after they are fixed, is a recommended practice so that attackers cannot exploit unpatched SVs to compromise systems~\cite{zhou2021finding}. One may argue that internal bug/SV reports in Issue Tracking Systems (ITS) such as JIRA~\cite{jira} or Bugzilla~\cite{bugzilla}
can be released before SV fixing and have severity levels. However, ITS severity levels are often for all bug types, not only SVs. These ITSs also do not readily provide exploitability and impact metrics like CVSS for SVs, limiting assessment information required for fixing prioritization. Moreover, SVs are mostly rooted in source code; thus, it is natural to perform code-based SV assessment.
We propose predicting seven base CVSS metrics after SVs are detected in code functions to enable thorough and prior-fixing SV assessment. We do not perform SV assessment for individual lines as for a given function, like Li et al.~\cite{li2021sysevr}, we observed that there can be more than one vulnerable line and nearly all these lines are strongly related and contribute to the same SV (having the same CVSS metrics).

Vulnerable statements represent the core parts of SVs, but we posit that other (non-vulnerable) parts of a function may also be usable for SV assessment.
Specifically, non-vulnerable statements in a vulnerable function are either \textit{directly} or \textit{indirectly} related to the current SV.
We use program slicing~\cite{weiser1984program} to define directly SV-related statements as the lines affect or are affected by the variables in vulnerable statements.
For example, the blue lines in \fig~\ref{fig:ex_sv} are directly related to the SV as they define, change, or use the \code{sb} and \code{dir} variables in vulnerable line 8. These SV-related statements can reveal the context/usage of affected variables for analyzing SV exploitability, impact, and severity. For instance, lines 5-6 denote that \code{dir} is a directory and \code{sb} is a string (\code{StringBuilder} object), respectively; line 7 then indicates that a directory change is performed, i.e., the \code{cd} command. This sequence of statements suggests that \code{sb} contains a command changing directory. Line 11 returns the vulnerable command, probably affecting other components.
Besides, indirectly SV-related statements, e.g., the black lines in \fig~\ref{fig:ex_sv}, are remaining lines in a function excluding vulnerable and directly SV-related statements. These indirectly SV-related lines may still provide information about SVs. For example, lines 3-4 in \fig~\ref{fig:ex_sv} imply that there is only a \code{null} checking for directory without imposing any privilege requirement to perform the command, potentially reducing the complexity of exploiting the SV. It remains unclear to what extent different types of statements are useful for SV assessment tasks. Therefore, this study aims to unveil the contributions of these statement types to function-level SV assessment models.

\section{Research Questions}
\label{sec:rqs}

To demystify the predictive performance of SV assessment models using vulnerable and other statements in code functions, we investigate the following three Research Questions (RQs).

\textbf{RQ1: Are vulnerable code statements more useful than non-vulnerable counterparts for SV assessment models?}
Since vulnerable and non-vulnerable statements are both potentially useful for SV assessment (see section~\mbox{\ref{subsec:vuln_functions}}), RQ1 compares them for building function-level SV assessment models.
RQ1 tests the hypothesis that vulnerable statements directly causing SVs would provide an advantage in SV assessment performance.
The findings of RQ1 would also inform the practice of leveraging recent advances in fine-grained SV detection for function-level SV assessment.

\textbf{RQ2: To what extent do different types of context of vulnerable statements contribute to SV assessment performance?}
RQ2 studies the impact of using directly and indirectly SV-related statements as context for vulnerable statements, as discussed in section~\ref{subsec:vuln_functions}, on the performance of SV assessment in functions. We compare the performance of models using different types of context lines (see section~\ref{subsec:code_context}) that have been commonly used in the literature. RQ2 findings would unveil what types of context (if any) would be beneficial to use alongside vulnerable statements for developing function-level SV assessment models.

\textbf{RQ3: Does separating vulnerable statements and context to provide explicit location of SVs improve assessment performance?}
For SV assessment, RQ2 combines vulnerable statements and their context as a single input following their order in functions, while RQ3 treats these two types of statements as two separate inputs. Separate inputs explicitly specify which statements are vulnerable in each function for assessment models. RQ3 results would give insights into the usefulness of the exact location of vulnerable statements for function-level SV assessment models.

\section{Research Methodology}
\label{sec:method}

This section presents the experimental setup we used to perform a large-scale study on function-level SV assessment to support prioritization of SVs before fixing. We used a computing cluster with 16 CPU cores and 16GB of RAM to conduct all the experiments.

\noindent \textbf{Workflow overview}. \fig~\ref{fig:workflow} presents the workflow we followed to develop function-level SV assessment models based on various types of code inputs. The workflow has three main steps: (\textit{i}) Collection of vulnerable and non-vulnerable statements from Vulnerability-Fixing Commits (VFCs) (section~\ref{subsec:data_collection}), (\textit{ii}) Context extraction of vulnerable statements (section~\ref{subsec:code_context}), and (\textit{iii}) Model building for SV assessment (sections~\ref{subsec:code_features},~\ref{subsec:assessment_models}, and~\ref{subsec:model_evaluation}).
We start with VFCs containing code changes used to fix SVs. As discussed in section~\ref{subsec:vuln_functions}, we consider the deleted (--) lines in each function of VFCs as vulnerable statements, while the remaining lines are non-vulnerable statements. Details of the extracted VFCs and statements are given in section~\ref{subsec:data_collection}. Both vulnerable and non-vulnerable statements are used by the \textit{Context Extraction} module (see section~\ref{subsec:code_context}) to obtain the three types of context with respect to vulnerable statements that potentially provide additional information for SV assessment. The extracted statements along with their context enter the \textit{Model Building} module. The first step in this module is to extract fixed-length feature vectors from code inputs/statements (see section~\ref{subsec:code_features}). Subsequently, such feature vectors are used to train different data-driven models (see section~\ref{subsec:assessment_models}) to support automated SV assessment, i.e., predicting the seven CVSS metrics: Access Vector, Access Complexity, Authentication, Confidentiality, Integrity, Availability, and Severity. The model training and evaluation are repeated 10 times to increase the stability of results (see section~\ref{subsec:model_evaluation}).

\noindent \textbf{RQ-wise method}. The methods to collect data, extract features as well as develop and evaluate models in \fig~\ref{fig:workflow} were utilized for answering all the Research Questions (RQs) in section~\ref{sec:rqs}.
\textbf{RQ1} developed and compared two types of SV assessment models, namely models using only vulnerable statements and those using only non-vulnerable statements.
In \textbf{RQ2}, for each of the program slicing, surrounding, and function context types, we created a single feature vector by combining the current context and corresponding vulnerable statements, based on their appearance order in the original functions, for model building and performance comparison. In \textbf{RQ3}, for each context type in RQ2, we extracted two separate feature vectors, one from vulnerable statements and another one from the context, and then fed these vectors into SV assessment models. We compared the two-input approach in RQ3 with the single-input counterpart in RQ2.

\subsection{Data Collection}
\label{subsec:data_collection}

To develop SV assessment models, we need a large dataset of vulnerable functions and statements curated from VFCs, as discussed in section~\ref{subsec:vuln_functions}. This section describes the collection of such dataset.

\noindent \textbf{VFC identification}. We first scraped VFCs from three popular sources in the literature: NVD~\mbox{\cite{nvd_website}}, GitHub Advisory Database,\footnote{https://github.com/advisories} and VulasDB~\mbox{\cite{ponta2019manually}}, a manually curated VFC dataset. The VFCs had dates ranging from July 2000 to September 2021. We only selected VFCs that had the CVSS version 2 metrics as we needed these metrics for SV assessment. Following the recommendation of~\mbox{\cite{mcintosh2017fix}}, we removed any VFCs that had more than 100 files and 10,000 lines of code as these VFCs are likely tangled commits, i.e., not only fixing SVs. We also discarded VFCs that were not written in the Java programming language as Java has been commonly used in both practice\footnote{https://bit.ly/stack-overflow-survey-2021} and the literature (e.g.,~\mbox{\cite{hoang2019deepjit,alon2019code2vec,mcintosh2017fix,le2021deepcva}}). After the filtering process, we obtained 900 VFCs to extract vulnerable functions/statements for building SV assessment models.

\begin{figure}[t]
    \centering
    \includegraphics[width=\columnwidth,keepaspectratio]{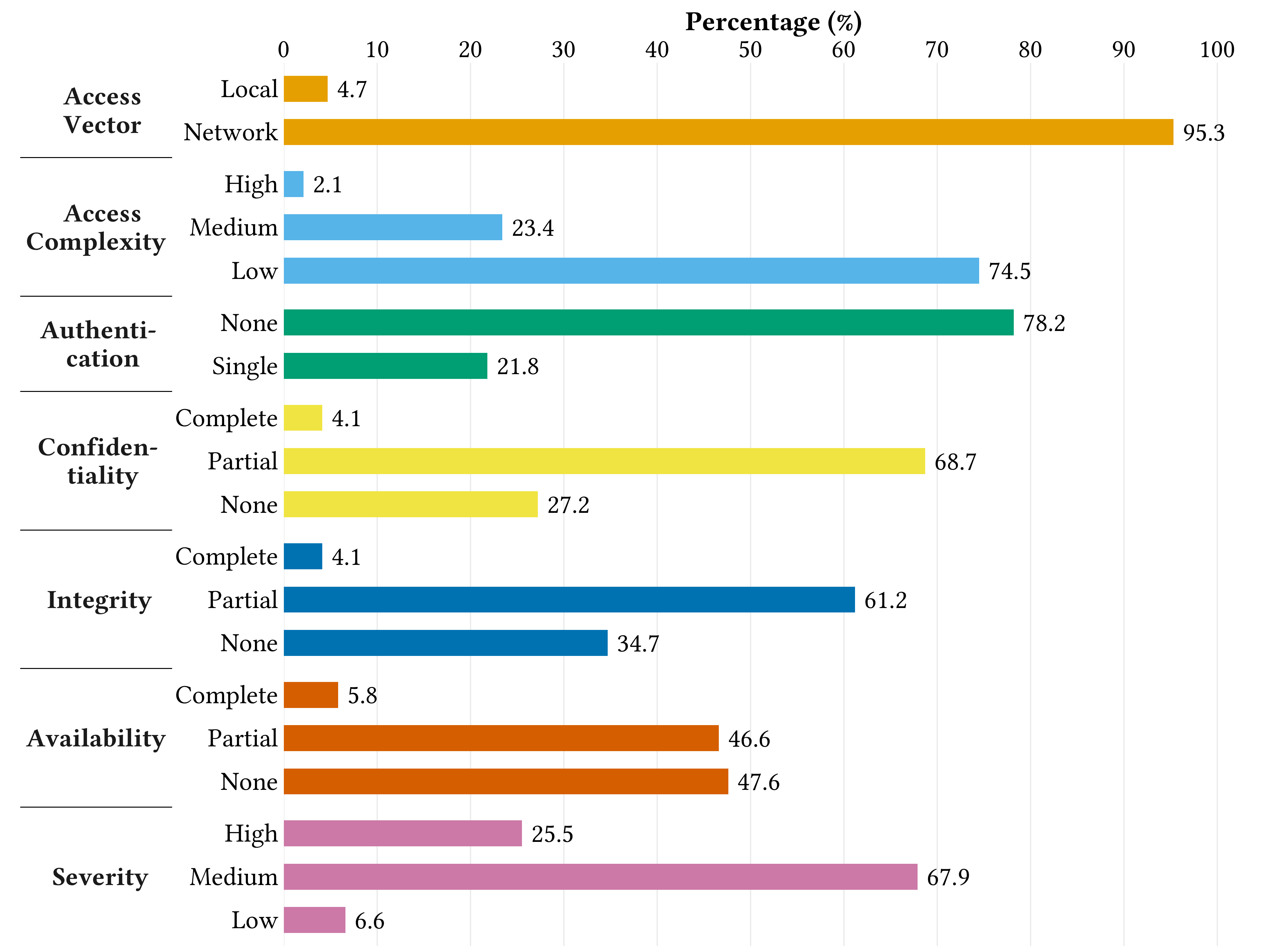}
    \caption{Class distributions of the seven CVSS metrics.}
    \label{fig:cvss_distribution}
\end{figure}

\noindent \textbf{Extraction of vulnerable functions and statements}. For each VFC, we obtained all the affected files (i.e., containing changed lines), excluding test files because we focused on production code. We followed a common practice~\cite{li2021vulnerability,wattanakriengkrai2020predicting,li2020dlfix} to consider all the functions in each affected file as \textit{vulnerable functions} and the deleted lines in these functions as \textit{vulnerable statements}. We removed functions having only added changes and non-functional/cosmetic changes such as removing/changing inline/multi-line comments, spaces, or empty lines. For the former, added lines only exist in fixed code, making it hard to pinpoint the exact vulnerable statements or root causes leading to such code additions~\cite{rezk2021ghost}. For the latter, cosmetic changes likely do not contribute to SV fixes~\cite{mcintosh2017fix}. We also did not use a function if its entire body was deleted because such a case did not have any non-vulnerable statements for building SV assessment models in our RQs (see section~\ref{sec:rqs}). After the filtering steps, we retrieved 1,782 vulnerable functions and 5,179 vulnerable statements of 429 SVs in 200 Java projects. We also obtained the seven CVSS metrics from NVD for each vulnerable function (see \fig~\ref{fig:cvss_distribution}).

\noindent \textbf{Manual validation of vulnerable functions}.
We randomly selected 317 functions, i.e., with 95\% confidence level and 5\% error~\mbox{\cite{cochran2007sampling}}, from our dataset. The first author and a PhD student with three-year experience in Software Engineering and Cybersecurity independently validated the functions.
The manual validation was considerably labor-intensive, taking 120 man-hours.
We achieved a substantial agreement with a Cohen's kappa score~\mbox{\cite{viera2005understanding}} of 0.72.
Disagreements were resolved through discussion.
We found that 9\% of the selected functions were not vulnerable, mainly due to tangled fixes/VFCs.
The functions in these VFCs fixed a non-SV related issue, e.g., the \code{nameContainsForbiddenSequence} function in the commit \textit{cefbb94} of the \textit{Eclipse Mojarra} project.
The modifier of this function in the \code{ResourceManager.java} class was changed from \code{private} to \code{package}. This change allowed the reuse of the function in other classes like \code{ClasspathResourceHelper.java} for sanitizing inputs to prevent a path traversal SV (CVE-2020-6950).
We assert that it is challenging to detect all of these cases without manual validation. However, such validation is extremely expensive to scale up to thousands of functions like the dataset we curated.

\subsection{Vulnerable Code Context Extraction}
\label{subsec:code_context}
This section describes the \textit{Context Extraction} module that takes vulnerable and non-vulnerable statements as inputs and then outputs program slicing, surrounding, or function context.
These context types have been previously used for bug/SV-related tasks~\cite{li2021sysevr,tian2020evaluating,lin2020deep}. However, there is little known about their use and value for function-level SV assessment tasks, which are studied in this work.

\noindent \textbf{Program slicing context}. Program slicing captures relevant statements to a point in a program (line of code) to support software debugging~\cite{weiser1984program}.
This concept has been utilized for SV identification~\cite{li2021sysevr,zheng2021vulspg}. However, using this context for SV assessment is fundamentally different. For SV detection, the location of vulnerable statements is unknown, so program slicing context is usually extracted for all statements including non-vulnerable ones in a function of pre-defined types (e.g., array manipulations, arithmetic operations, and function calls)~\cite{li2021sysevr}. In contrast, for SV assessment, vulnerable statements are known; thus, program slicing context is only obtained for these statements.
With a focus on function-level SV assessment, we considered \textit{intra-procedural program slicing}, i.e., finding relevant lines within the boundary of a function of interest.

Following the common practice in the literature~\cite{salimi2020improving,li2021sysevr}, we used Program Dependence Graph (PDG)~\cite{ferrante1987program} extracted from source code to obtain program slices for vulnerable statements in each function. A PDG contains nodes that represent code statements and directed edges that capture data or control dependencies among nodes. A data dependency exists between two statements when one statement affects/changes the value of a variable used in another statement. For example, ``\code{int b = a + 1;}'' is data-dependent on ``\code{int a = 1;}'' as the variable \code{a} defined in the second statement is used in the first statement. A control dependency occurs when a statement determines whether/how often another statement is executed. For instance, in ``\code{if (b == 2) func(c);}'', ``\code{func(c)}'' only runs if ``\code{b == 2}'', and thus is control-dependent on the former.

Based on data and control dependencies, backward and forward slices are extracted.
\textit{Backward slices} directly change or control the execution of statements affecting the values of variables in vulnerable statements; whereas, \textit{forward slices} are data/control-dependent on vulnerable statements~\cite{dashevskyi2018screening}.
In a PDG, backward slices are nodes that can go to vulnerable nodes through one or more directed edges. In \fig~\ref{fig:ex_sv}, the \code{dir} variable is defined in line 5 and then used in vulnerable line 8, so line 5 is a backward slice. Forward slices are the nodes that can be reached from vulnerable nodes by following one or more directed edges in a PDG. In \fig~\ref{fig:ex_sv}, line 11 is data-dependent on vulnerable line 8 as it uses the value of \code{sb}; thus, line 11 is a forward slice. The program slicing context of vulnerable statements in a function is a combination of all backward and forward slices.

\noindent \textbf{Surrounding context}. Another way to define context is to take a fixed number of lines ($n$) before and after a vulnerable statement, which is referred to as \textit{surrounding context}. These surrounding lines may contain relevant information, e.g., values/usage of variables in vulnerable statements. This context is also based on an observation that developers usually first look at nearby code of vulnerable statements to understand how to fix SVs~\cite{tian2020evaluating}.
We discarded surrounding lines that were just code comments or blank lines as these probably do not contribute to the functionality of a function~\cite{mcintosh2017fix}. We also limited surrounding lines to be within-function.

\noindent \textbf{Function context}. Contrary to program slicing and surrounding context that may not use all lines in a function, \textit{function context} uses all function statements, excluding vulnerable ones. This scope has been commonly used for SV detection models~\cite{zhou2019devign,lin2020deep} because vulnerable statements are unavailable at detection time. This scope contains all the lines of program slicing/surrounding context and other presumably indirectly related lines to vulnerable statements. Accordingly, the performance of using indirectly SV-related lines together with directly SV-relevant lines for SV assessment can be examined.
Note that for a given function, combining function context with vulnerable statements as a single code block (RQ2 in section~\ref{sec:rqs}) is equivalent to using the whole function, which would result in the same input to SV assessment models regardless of which statements are vulnerable.
This input combination allows us to evaluate the usefulness of the exact location of vulnerable statements for function-level SV assessment models.

\subsection{Code Feature Generation}
\label{subsec:code_features}

Raw code from vulnerable statements and their context are converted into fixed-length feature vectors to be consumable by learning-based SV assessment models. This step describes five techniques we used to extract features from code inputs.

\noindent \textbf{Bag-of-Tokens}. \textit{Bag-of-Tokens} is based on Bag-of-Words, a popular feature extraction technique in Natural Language Processing (NLP). This technique has been commonly investigated for developing SV assessment models based on textual SV descriptions/reports~\cite{wen2015novel,gawron2017automatic,wang2019intelligent}. We extended this technique to code-based SV assessment by counting the frequency of code tokens. We also applied code-aware tokenization to preserve code syntax and semantics. For instance, \code{var++} was tokenized into \code{var} and \code{++}, explicitly informing a model about incrementing the variable \code{var} by one using the operator (\code{++}).

\noindent \textbf{Bag-of-Subtokens}. \textit{Bag-of-Subtokens} extends Bag-of-Tokens by splitting extracted code tokens into sequences of characters (sub-tokens). These characters help a model learn less frequent tokens better. For instance, an infrequent variable like \code{ThisIsAVeryLongVar} is decomposed into multiple sub-tokens; one of which is \code{Var}, telling a model that this is a potential variable. We extracted sub-tokens of lengths ranging from two to six. Such values have been previously adopted for SV assessment~\cite{nakagawa2019character,le2019automated}. We did not use one-letter characters as they were too noisy, while using more than six characters would significantly increase feature size and computational cost.

\noindent \textbf{Word2vec}. Unlike Bag-of-Tokens and Bag-of-Subtokens that do not consider token context, \textit{Word2vec}~\cite{mikolov2013distributed} extracts features of a token based on its surrounding counterparts. The contextual information from surrounding tokens helps produce similar feature vectors in an embedding space for tokens with (nearly) identical functionality/usage (e.g., \code{average} and \code{mean} variables). Word2vec generates vectors for individual tokens, so we averaged the vectors of all input tokens to represent a code snippet. This averaging method has been demonstrated to be effective~\cite{shen2018baseline}.
\tab~\ref{tab:hyperparameter_models} lists different values for the window and vector sizes of Word2vec used for tuning the performance of learning-based SV assessment models.

\noindent \textbf{fastText}. \textit{fastText}~\cite{bojanowski2017enriching} enhances Word2vec by representing each token with an aggregated feature vector of its constituent sub-tokens. Technically, fastText combines the strengths of semantic representation of Word2vec and subtoken-augmented features of Bag-of-Subtokens. fastText has been shown to build competitive yet compact report-level SV assessment models~\cite{le2019automated}. Like Word2vec, the feature vector of a code snippet was averaged from the vectors of all the input tokens. The length of sub-tokens also ranged from two to six, resembling that of Bag-of-Subtokens.
Other hyperparameters of fastText for optimization are listed in \tab~\ref{tab:hyperparameter_models}.

\noindent \textbf{CodeBERT}. \textit{CodeBERT}~\cite{feng2020codebert} is an adaptation of BERT~\cite{devlin2018bert}, the current state-of-the-art feature representation technique in NLP, to source code modeling. CodeBERT is a pre-trained model using both natural language and programming language data to produce contextual embedding for code tokens.
The same code token can have different CodeBERT embedding vectors depending on other tokens in an input; whereas, word2vec/fastText produces a single vector for every token regardless of its context. In addition, the source code tokenizer of CodeBERT is built upon Byte-Pair Encoding (BPE)~\cite{sennrich2015neural}. This tokenizer smartly retains sub-tokens that frequently appear in a training corpus rather than keeping all of them as in Bag-of-Subtokens and fastText, balancing between performance and cost. CodeBERT also preserves a special token, \textit{[CLS]}, to represent an entire code input. We leveraged the vector of this \textit{[CLS]} token to extract the features for each code snippet.

We trained all the feature models from scratch, except CodeBERT as it is a pre-trained model. We used CodeBERT's pre-trained vocabulary and embeddings as commonly done in the literature~\cite{zhou2021assessing}. To build the vocabulary for the other feature extraction methods, we considered tokens appearing at least in two samples in a training dataset to avoid vocabulary explosion due to too rare tokens. The exact vocabulary depended on the dataset used in each of the 10 training/evaluation rounds, as described in section~\ref{subsec:model_evaluation}.
Note that some code snippets, e.g., vulnerable lines extracted from (partial) code changes, were not compilable (i.e., did not contain complete code syntax); thus, we did not use Abstract Syntax Tree (AST) based code representation like Code2vec~\cite{alon2019code2vec} in this study as such representation may not be robust for these cases~\cite{hoang2019deepjit,hoang2020cc2vec}. It is worth noting that Bag-of-Tokens, Bag-of-Subtokens, Word2vec, fastText, and CodeBERT can still work with these cases as these methods operate directly on code tokens.

\subsection{Data-driven SV Assessment Models}
\label{subsec:assessment_models}

\begin{table}[t]
\fontsize{8}{9}\selectfont
  \centering
  \caption{Hyperparameter tuning for SV assessment models.}
    \begin{tabular}{lll}
    \hline
    \multicolumn{1}{l}{\textbf{Step}} & \multicolumn{1}{l}{\textbf{Model}} & \multicolumn{1}{l}{\textbf{Hyperparameters}} \\
    \hline
    \multirow{2}{*}{\makecell[l]{Feature\\ extraction}} & \makecell[l]{Word2vec~\cite{shen2018baseline}} & \makecell[l]{\textit{Vector size}: 150, 300, 500} \\
    & \makecell[l]{fastText~\cite{bojanowski2017enriching}} & \makecell[l]{\textit{Window size}: 3, 4, 5} \\
    \hline
    \multirow{8}{*}{\makecell[l]{CVSS\\ metrics\\ prediction}} & \makecell[l]{Logistic Regression\\ (LR)~\cite{walker1967estimation}} & \multirow{3}{*}{\makecell[l]{\textit{Regularization coefficient}:\\ 0.01, 0.1, 1, 10, 100}} \\
    & \makecell[l]{Support Vector\\ Machine (SVM)~\cite{cortes1995support}} & \\
    \cline{2-3}
     & \multirow{3}{*}{\makecell[l]{K-Nearest\\ Neighbors\\ (KNN)~\cite{altman1992introduction}}} & \makecell[l]{\textit{No. of neighbors}: 5, 11, 31, 51} \\
    & & \makecell[l]{\textit{Weight}: uniform, distance} \\
    & & \textit{Distance norm}: 1, 2 \\
    \cline{2-3}
    & \makecell[l]{Random Forest (RF)~\cite{ho1995random}} & \multirow{3}{*}{\makecell[l]{\textit{No. of estimators}: 100, 200,\\ 300, 400, 500\\ \textit{Max depth}: 3, 5, 7, 9,\\ unlimited\\ \textit{Max. no. of leaf nodes}: 100,\\ 200, 300, unlimited (RF)}} \\
    & \makecell[l]{Extreme Gradient\\ Boosting (XGB)~\cite{chen2016xgboost}} & \\
    & \makecell[l]{Light Gradient\\ Boosting Machine\\ (LGBM)~\cite{ke2017lightgbm}} & \\
    \hline
    \end{tabular}%
  \label{tab:hyperparameter_models}%
\end{table}%

Features generated from code inputs enter ML models for predicting the CVSS metrics. The predictions of the CVSS metrics are classification problems (see \fig~\ref{fig:cvss_distribution}). We used six well-known Machine Learning (ML) models for classifying the classes of each CVSS metric: Logistic Regression (LR)~\cite{walker1967estimation}, Support Vector Machine (SVM)~\cite{cortes1995support}, K-Nearest Neighbors (KNN)~\cite{altman1992introduction}, Random Forest (RF)~\cite{ho1995random}, eXtreme Gradient Boosting (XGB)~\mbox{\cite{chen2016xgboost}}, and Light Gradient Boosting Machine (LGBM)~\mbox{\cite{ke2017lightgbm}}. LR, SVM, and KNN are single models, while RF, XGB, and LGBM are ensemble models that leverage multiple single counterparts to reduce overfitting. These classifiers have been used for SV assessment based on SV reports~\mbox{\cite{spanos2018multi,le2019automated}}. We also considered different hyperparameters for tuning the performance of the classifiers, as given in \tab~\ref{tab:hyperparameter_models}. These hyperparameters have been adapted from the prior studies using similar classifiers~\mbox{\cite{spanos2018multi,le2019automated,le2020puminer}}. Here, we mainly focus on ML techniques, and thus using Deep Learning models~\cite{le2020deep} for the tasks is out of the scope of this study.

\subsection{Model Evaluation}
\label{subsec:model_evaluation}
\noindent \textbf{Evaluation technique}.
To develop function-level SV assessment models and evaluate their performance, we used 10 rounds of training, validation, and testing. We randomly shuffled the dataset of vulnerable functions in section~\ref{subsec:data_collection} and then split it into 10 partitions of roughly equal size.\footnote{With 1,782 samples in total, folds 1-9 had 178 samples and fold 10 had 180 samples.} In round $i$, for a model, we used fold $i + 1$ for validation, fold $i + 2$ for testing, and all of the remaining folds for training.
When $i + 1$ or $i + 2$ was larger than 10, its value was wrapped around. For example, if $i = 10$, then $(i + 1)~\text{mod}~10 = 11~\text{mod}~10 = 1$ and $(i + 2)~\text{mod}~10 = 12~\text{mod}~10 = 2$. A grid search of the hyperparameters in \tab~\ref{tab:hyperparameter_models} was performed using the validation sets to select optimal models. The performance of such optimal models on the test sets was reported.
It is important to note that our evaluation strategy improves upon 10-fold cross-validation and random splitting data into a single training/validation/test set, the two most commonly used evaluation techniques in (fine-grained) SV detection and report-level SV assessment studies~\cite{hanif2021rise,sonnekalb2022deep,li2021vuldeelocator,li2021vulnerability,nguyen2021information,le2021survey}. Our evaluation technique has separate test sets, which cross-validation does not, to objectively measure the performance of tuned/optimal models on unseen data. Using multiple (10) validation/test sets also increases the stability of results compared to a single set~\cite{raschka2018model}.
Moreover, we aim to provide baseline performance for function-level SV assessment in this study, so we did not apply any techniques like class rebalancing or feature selection/reduction to augment the data/features. Such augmentation can be explored in future work.
We also did not compare SV assessment using functions with that using reports as their use cases are different; function-level SV assessment is needed when SV reports are unavailable/unusable. It may be fruitful to compare/combine these two artifacts for SV assessment in the future.

\noindent \textbf{Evaluation measures}.
We used F1-Score\footnote{The macro version of F1-Score was used for multi-class classification.} and Matthews Correlation Coefficient (MCC) measures to quantify how well developed models perform SV assessment tasks. F1-Score values are from 0 to 1, and MCC has values from –1 to 1; 1 is the best value for both measures. These measures have been commonly used for SV assessment (e.g.,~\mbox{\cite{spanos2018multi,kudjo2019improving,le2019automated}}).
We used MCC as the main measure for selecting optimal models because MCC takes into account all classes, i.e., all cells in a confusion matrix, during evaluation~\mbox{\cite{luque2019impact}}.

\noindent \textbf{Statistical analysis}. To confirm the significance of results, we employed one-sided Wilcoxon signed rank test~\cite{wilcoxon1992individual} and its respective effect size ($r = Z / \sqrt{N}$, where $Z$ is the $Z$-score statistic of the test and $N$ is the total count of samples)~\cite{tomczak2014need}.\footnote{$r \leq 0.1$: negligible, $0.1 < r \leq 0.3$: small, $0.3 < r \leq 0.5$: medium, $r > 0.5$: large~\cite{field2013discovering}} We used Wilcoxon signed-rank test because it is a non-parametric test that can compare two-paired groups of data, and we considered a test significant if its confidence level was more than 99\% ($p$-value $<$ 0.01).
We did not use the popular Cohen's D~\cite{cohen2013statistical} and Cliff's $\delta$~\cite{macbeth2011cliff} effect sizes as they are not suitable for comparing paired data~\cite{wattanakriengkrai2020predicting}.

\vspace{-3pt}
\section{Results}
\label{sec:results}

\subsection{RQ1: Are vulnerable code statements more useful than non-vulnerable counterparts for SV assessment models?}
\label{subsec:rq1_results}

\begin{figure}[t]
    \centering
    \includegraphics[width=\columnwidth,keepaspectratio]{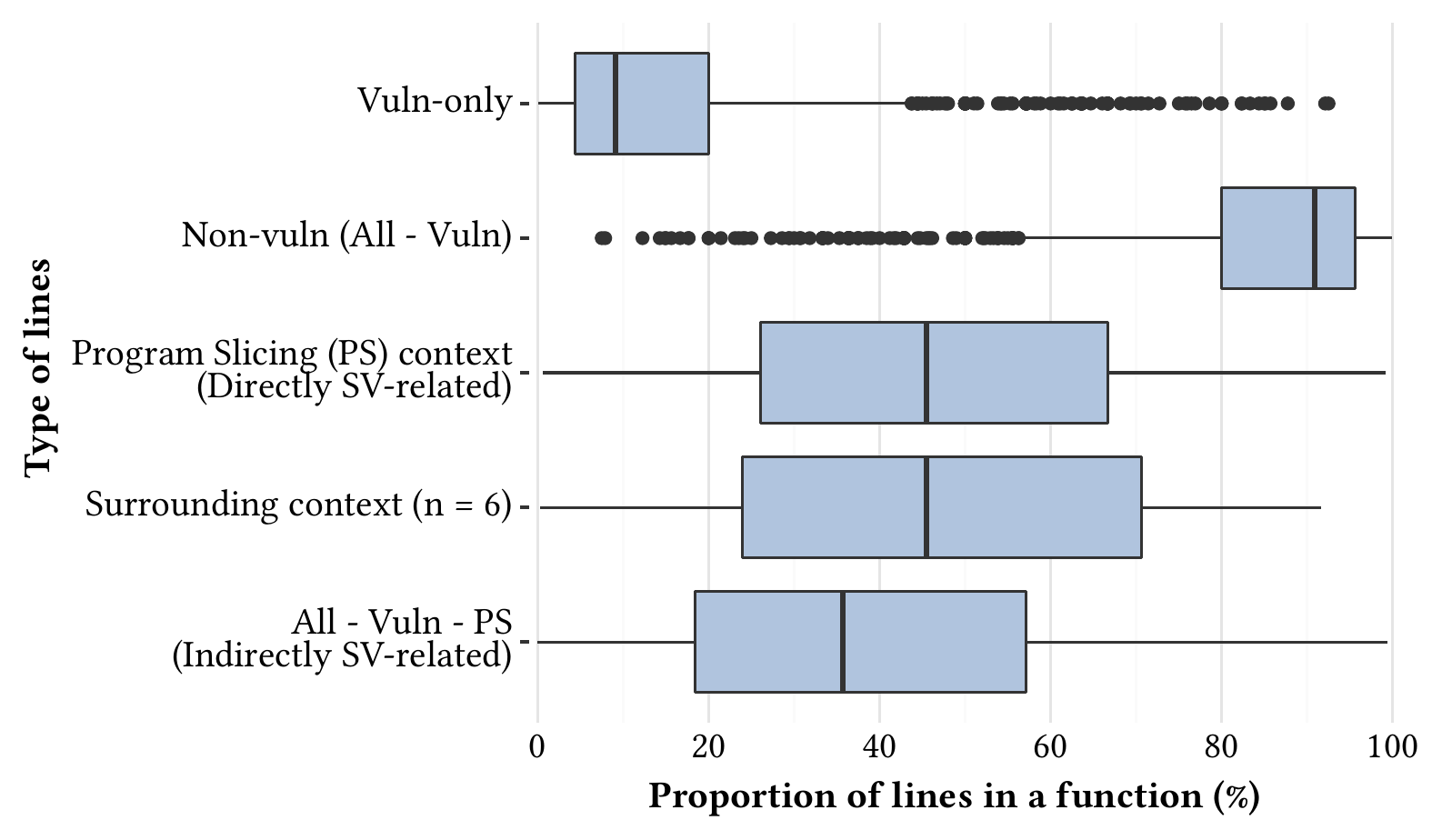}
    \caption{Proportions of different types of lines in a function. Notes: Min proportion of Vuln-only lines is non-zero (0.03\%); thus, max of Non-vuln line proportion is $<$ 100\%. Cosmetic lines were excluded from the computation of ratios.}
    \label{fig:scope_proportion}
\end{figure}

Based on the extraction process in section~\ref{subsec:data_collection}, we collected 1,782 vulnerable functions containing 5,179 vulnerable and 57,633 non-vulnerable statements. The proportions of these two types of statements are given in the first and second boxplots, respectively, in \fig~\ref{fig:scope_proportion}. On average, 14.7\% of the lines in the selected functions were vulnerable, 5.8 times smaller than that of non-vulnerable lines.
Interestingly, we also observed that 55\% of the functions contained only a single vulnerable statement. These values show that vulnerable statements constitute a very small proportion of functions.

\textbf{\textit{Despite the small size (no. of lines), vulnerable statements outperformed non-vulnerable statements for the seven assessment tasks (see \tab~\ref{tab:vuln_vs_nonvuln})}}.
We considered two variants of non-vulnerable statements for comparison. The first variant, \textit{Non-vuln (random)}, randomly selected the same number of lines as vulnerable statements from non-vulnerable statements in each function.
The second variant, \textit{Non-vuln (all)} aka. Non-vuln (All - Vuln) in \fig~\ref{fig:scope_proportion}, considered all non-vulnerable statements. Compared to same-sized non-vulnerable statements (Non-vuln (random)), \textit{Vuln-only} (using vulnerable statements solely) produced 116.9\%, 126.6\%, 98.7\%, 90.7\%, 147.9\%, 111.2\%, 116.7\% higher MCC for Access Vector, Access Complexity, Authentication, Confidentiality, Integrity, Availability, and Severity tasks, respectively. On average, Vuln-only was 114.5\% and 43.6\% better than Non-vuln (random) in MCC and F1-Score, respectively. We obtained similar results of Non-vuln (random) when repeating the experiment with differently randomized lines. When using all non-vulnerable statements (Non-vuln (all)), the assessment performance increased significantly, yet was still lower than that of vulnerable statements.
Average MCC and F1-Score of Vuln-only were 7.4\% and 5.5\% higher than Non-vuln (all), respectively.
The improvements of Vuln-only over the two variants of non-vulnerable statements were statistically significant across features/classifiers with $p$-values $<$ 0.01 ($p$-value$_{Non-vuln (random)} = 1.7 \times 10^{-36}$ and $p$-value$_{Non-vuln (all)} = 7.2 \times 10^{-11}$) and non-negligible effect sizes ($r_{Non-vuln (random)} = 0.62$ and $r_{Non-vuln (all)} = 0.32$).
The low performance of Non-vuln (random) implies that SV assessment models likely perform worse if vulnerable statements are incorrectly identified. Moreover, the decent performance of Non-vuln (all) shows that some non-vulnerable statements are potentially helpful for SV assessment, which are studied in detail in RQ2.

\begin{table}[t]
\fontsize{6.9}{7.9}\selectfont
  \centering
  \caption{Testing performance for SV assessment tasks of vulnerable vs. non-vulnerable statements. Note: Bold and grey-shaded values are best row-wise performance.}
    \begin{tabular}{llccc}
    \hline
    \multirowcell{3}[0ex][l]{\textbf{CVSS metric}} & \multirowcell{3}[0ex][l]{\textbf{Evaluation}\\ \textbf{metric}} & \multicolumn{3}{c}{\textbf{Input type}}\\
    \cline{3-5}
    & & \textbf{Vuln-only} & \makecell{\textbf{Non-vuln}\\ \textbf{(random)}} & \makecell{\textbf{Non-vuln}\\ \textbf{(all)}} \\
    \hline
    \multirowcell{2}[0ex][l]{Access Vector} & F1-Score & \cellcolor[HTML]{C0C0C0} \textbf{0.820} & 0.650 & 0.786 \\
    & MCC & \cellcolor[HTML]{C0C0C0} \textbf{0.681} & \makecell{0.314} & \makecell{0.605}\\
    \hline
    \multirowcell{2}[0ex][l]{Access\\ Complexity} & F1-Score & \cellcolor[HTML]{C0C0C0} \textbf{0.622} & 0.458 & 0.592 \\
    & MCC & \cellcolor[HTML]{C0C0C0} \textbf{0.510} & \makecell{0.225} & \makecell{0.467}\\
    \hline
    \multirowcell{2}[0ex][l]{Authentication} & F1-Score & \cellcolor[HTML]{C0C0C0} \textbf{0.791} & 0.602 & 0.765 \\
    & MCC & \cellcolor[HTML]{C0C0C0} \textbf{0.630} & \makecell{0.317} & \makecell{0.614}\\
    \hline
    \multirowcell{2}[0ex][l]{Confidentiality} & F1-Score & \cellcolor[HTML]{C0C0C0} \textbf{0.645} & 0.411 & 0.625 \\
    & MCC & \cellcolor[HTML]{C0C0C0} \textbf{0.574} & \makecell{0.301} & 0.561\\
    \hline
    \multirowcell{2}[0ex][l]{Integrity} & F1-Score & \cellcolor[HTML]{C0C0C0} \textbf{0.650} & 0.384 & 0.616 \\
    & MCC & \cellcolor[HTML]{C0C0C0} \textbf{0.585} & \makecell{0.236} & \makecell{0.534}\\
    \hline
    \multirowcell{2}[0ex][l]{Availability} & F1-Score & \cellcolor[HTML]{C0C0C0} \textbf{0.647} & 0.417 & 0.624 \\
    & MCC & \cellcolor[HTML]{C0C0C0} \textbf{0.583} & \makecell{0.276} & \makecell{0.551}\\
    \hline
    \multirowcell{2}[0ex][l]{Severity} & F1-Score & \cellcolor[HTML]{C0C0C0} \textbf{0.695} & 0.414 & 0.610 \\
    & MCC & \cellcolor[HTML]{C0C0C0} \textbf{0.583} & \makecell{0.269} & \makecell{0.523}\\
    \hline
    \hline
    \multirowcell{2}[0ex][l]{Average} & F1-Score & \cellcolor[HTML]{C0C0C0} \textbf{0.695} & 0.484 & 0.659\\
    & MCC & \cellcolor[HTML]{C0C0C0} \textbf{0.592} & 0.276 & 0.551 \\
    \hline
    \end{tabular}%
  \label{tab:vuln_vs_nonvuln}%
\end{table}%

\subsection{RQ2: To what extent do different types of context of vulnerable statements contribute to SV assessment performance?}
\label{subsec:rq2_results}

\begin{figure*}[t]
    \begin{subfigure}{\textwidth}
         \centering
         \includegraphics[width=\columnwidth,keepaspectratio]{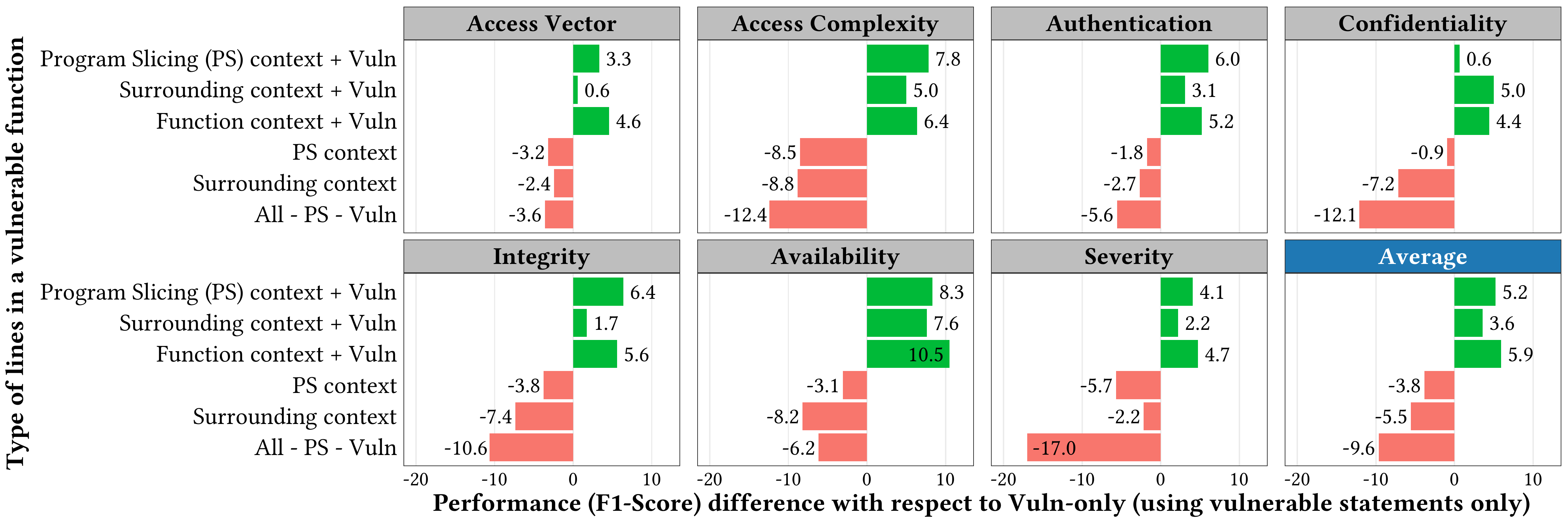}
         \label{fig:f1_diff}
     \end{subfigure}
     \rule{\textwidth}{0.01pt}
     \begin{subfigure}{\textwidth}
         \centering
         \includegraphics[width=\columnwidth,keepaspectratio]{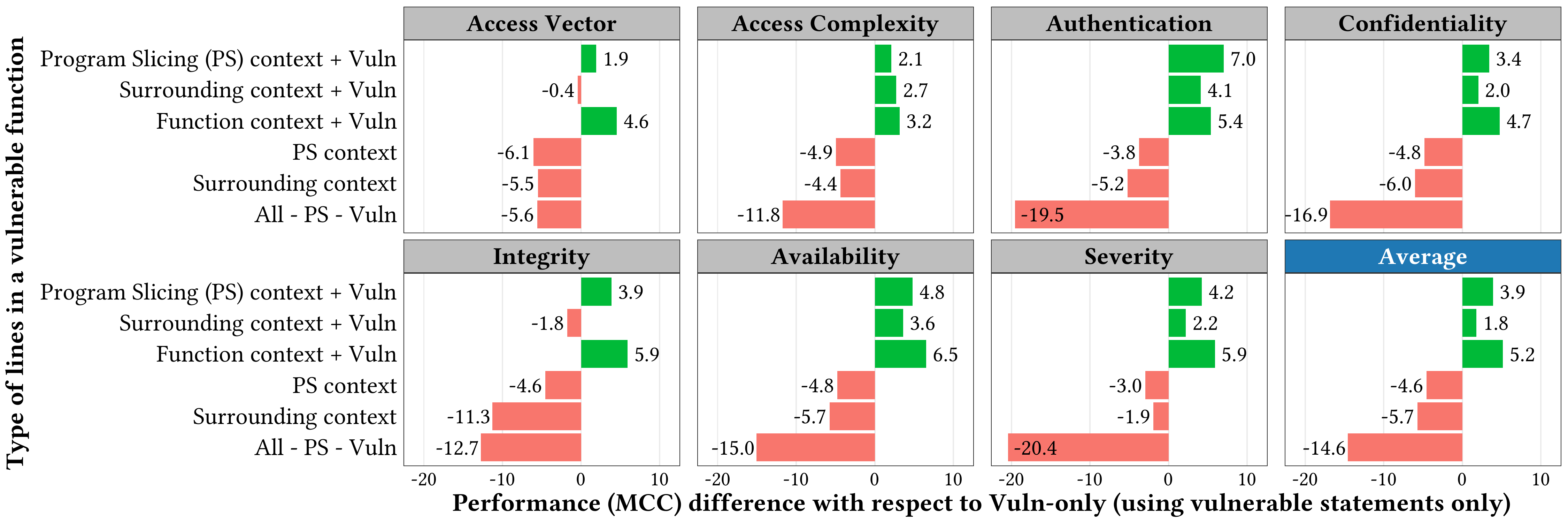}
         \label{fig:mcc_diff}
     \end{subfigure}
     \caption{Differences in testing SV assessment performance (F1-Score and MCC) between models using different types of lines/context and those using only vulnerable statements. Note: The differences were multiplied by 100 to improve readability.}
    \label{fig:context_comparison}
\end{figure*}

In RQ2, we compared the performance of models using \textit{Program Slicing} (PS), \textit{surrounding} and \textit{function} context. Notably, we removed 365 cases for which we could not extract PS context from the dataset curated in section~\ref{subsec:data_collection}. Apparently, these cases were the same as using only vulnerable statements, which would make comparisons biased, especially against Vuln-only itself. The vulnerable statements in these cases mainly did not contain any variable, e.g., using an unsafe function without any parameter.\footnote{https://bit.ly/3pg36mp} In this new dataset, PS and surrounding context approximately constituted 46\%, on average, of lines in vulnerable functions (see \fig~\ref{fig:scope_proportion}). We used six lines before and after vulnerable statements (n = 6) as surrounding context because this value resulted in the closest average context size to that of PS context. It is impossible to have the exactly same size because PS context is dynamically derived from vulnerable statements, while surrounding context is predefined. The roughly similar size helps test whether directly SV-related lines in PS context would be better than pre-defined surrounding lines for SV assessment.
The training and evaluation processes on the dataset in RQ2 were the same as in RQ1.\footnote{The RQ1 findings still hold when using the new dataset in RQ2, yet with a slight ($\approx$2\%) decrease in absolute model performance.}

\textit{\textbf{Overall, adding context to vulnerable statements led to better SV assessment performance than using vulnerable statements only (see \fig~\ref{fig:context_comparison}). Among the three, function context was the best, followed by PS and then surrounding context}}. In terms of the main measure MCC, function context working together with vulnerable statements beat Vuln-only by 6.4\%, 6.5\%, 9\%, 8.2\%, 11\%, 11.4\%, 9.7\% for Access Vector, Access Complexity, Authentication, Confidentiality, Integrity, Availability, and Severity tasks, respectively. The higher F1-Score values when incorporating function context to vulnerable statements are also evident in \fig~\ref{fig:context_comparison}.
On average, combining function context and vulnerable statements attained 0.64 MCC and 0.75 F1-Score, surpassing using vulnerable lines solely by 8.9\% in MCC and 8.5\% in F1-Score.
Although PS context + Vuln performed slightly worse than function context + Vuln, MCC and F1-Score of PS context + Vuln were still 6.7\% and 7.5\% ahead of Vuln-only, respectively. The improvements of function and PS context + Vuln over Vuln-only were significant across features/classifiers, i.e., $p$-values of $1.2 \times 10^{-17}$ and $2.1 \times 10^{-13}$ and medium effect sizes of 0.42 and 0.36, respectively. Compared to function/PS context + Vuln, surrounding context + Vuln outperformed Vuln-only by smaller margins, i.e., 3\% for MCC and 5.2\% for F1-Score ($p$-value = $3.7 \times 10^{-8} < 0.01$ with a small effect size ($r = 0.27$)). These findings show the usefulness of directly SV-related (PS) lines for SV assessment models, while six lines surrounding vulnerable statements seemingly contain less related information for the SV assessment tasks. Further investigation revealed that only 49\% of lines in PS context overlapped with those in surrounding context (n = 6). Note that the performance of surrounding context tended to approach that of function context as the surrounding context size increased. Using the dataset in RQ1, we also obtained the same patterns, i.e., function context + Vuln > surrounding context + Vuln > Vuln-only.
This result shows that function context is generally better than the other context types, indicating the plausibility of building effective SV assessment models using only the output of function-level SV detection (i.e., requiring no knowledge about which statements are vulnerable in each function).

\textit{\textbf{Although the three context types were useful for SV assessment when combined with vulnerable statements, using these context types alone significantly reduced the performance}}. As shown from RQ1, using only function context (i.e., Non-vuln (all) in \tab~\ref{tab:vuln_vs_nonvuln}) was 6.9\% inferior in MCC and 5.2\% lower in F1-Score than Vuln-only. Using the new dataset in RQ2, we obtained similar reductions in MCC and F1-Score values. \fig~\ref{fig:context_comparison} also indicates that using only PS and surrounding context decreased MCC and F1-Score of all the tasks. Particularly, using PS context alone reduced MCC and F1-Score by 7.8\% and 5.5\%, respectively; whereas, such reductions in values for using only surrounding context were 9.8\% and 8\%. These performance drops were confirmed significant with $p$-values < 0.01 and non-negligible effect sizes. Overall, the performance rankings of the context types with and without vulnerable statements were the same, i.e., function > PS > surrounding context.
We also observed that all context types were better (increasing 20.1-28.2\% in MCC and 6.9-17.5\% in F1-Score) than non-directly SV-related lines (i.e., All - PS - Vuln in \fig~\ref{fig:context_comparison}). These findings highlight the need for using context together with vulnerable statements rather than using each of them alone for function-level SV assessment tasks.

\begin{table}[t]
\fontsize{6.9}{7.9}\selectfont
  \centering
  \caption{Differences in testing performance for SV assessment tasks between double-input models (RQ3) and single-input models (RQ1/2). Notes: The differences were multiplied by 100 to increase readability. Green and red colors denote value increase and decrease, respectively. Darker color shows a higher magnitude of increase/decrease. Average performance values (multiplied by 100) of double-input models for the three context (ctx) types are in parentheses.
  }
    \begin{tabular}{llP{1.4cm}P{1.4cm}P{1.4cm}}
    \hline
    \multirowcell{3}[0ex][l]{\textbf{CVSS metric}} & \multirowcell{3}[0ex][l]{\textbf{Evaluation}\\ \textbf{metric}} & \multicolumn{3}{c}{\textbf{Input type (double)}}\\
    \cline{3-5}
    & & \makecell{\textbf{PS}\\ \textbf{ctx + Vuln}} & \makecell{\textbf{Surrounding}\\ \textbf{ctx + Vuln}} & \makecell{\textbf{Function}\\ \textbf{ctx + Vuln}} \\
    \hline
    \multirowcell{2}[0ex][l]{Access\\ Vector} & F1-Score & \cellcolor[HTML]{67cf82} 1.3 & \cellcolor[HTML]{6fd187} 1.2 & \cellcolor[HTML]{6fd187} 1.2 \\
    & MCC & \cellcolor[HTML]{07bc3d} 2.6 & \cellcolor[HTML]{00ba38} 2.7 & \cellcolor[HTML]{16bf48} 2.4\\
    \hline
    \multirowcell{2}[0ex][l]{Access\\ Complexity} & F1-Score & \cellcolor[HTML]{ff8a8a} -0.5 & \cellcolor[HTML]{9bdaa7} 0.6 & \cellcolor[HTML]{ff6e6e} -0.7 \\
    & MCC & \cellcolor[HTML]{58cc78} 1.5 & \cellcolor[HTML]{a2dbad} 0.5 & \cellcolor[HTML]{16bf48} 2.4\\
    \hline
    \multirowcell{2}[0ex][l]{Authentica-\\ tion} & F1-Score & \cellcolor[HTML]{a2dbad} 0.5 & \cellcolor[HTML]{93d8a2} 0.7 & \cellcolor[HTML]{b1deb7} 0.3 \\
    & MCC & \cellcolor[HTML]{7dd492} 1.0 & \cellcolor[HTML]{4ac96d} 1.7 & \cellcolor[HTML]{93d8a2} 0.7\\
    \hline
    \multirowcell{2}[0ex][l]{Confident-\\ iality} & F1-Score & \cellcolor[HTML]{ffb3b3} -0.2 & \cellcolor[HTML]{c6e3c6} 0.01 & \cellcolor[HTML]{ff5353} -0.9 \\
    & MCC & \cellcolor[HTML]{ff6e6e} -0.7 & \cellcolor[HTML]{ff8a8a} -0.5 & \cellcolor[HTML]{ff0000} -1.5\\
    \hline
    \multirowcell{2}[0ex][l]{Integrity} & F1-Score & \cellcolor[HTML]{b8e0bc} 0.2 & \cellcolor[HTML]{93d8a2} 0.7 & \cellcolor[HTML]{b8e0bc} 0.2 \\
    & MCC & \cellcolor[HTML]{ffc5c5} -0.07 & \cellcolor[HTML]{76d28d} 1.1 & \cellcolor[HTML]{a2dbad} 0.5\\
    \hline
    \multirowcell{2}[0ex][l]{Availability} & F1-Score & \cellcolor[HTML]{ff8a8a} -0.5 & \cellcolor[HTML]{ffb3b3} -0.2 & \cellcolor[HTML]{ff4545} -1.0\\
    & MCC & \cellcolor[HTML]{aaddb2} 0.4 & \cellcolor[HTML]{aaddb2} 0.4 & \cellcolor[HTML]{c0e1c2} 0.1\\
    \hline
    \multirowcell{2}[0ex][l]{Severity} & F1-Score & \cellcolor[HTML]{93d8a2} 0.7 & \cellcolor[HTML]{85d597} 0.9 & \cellcolor[HTML]{93d8a2} 0.7\\
    & MCC & \cellcolor[HTML]{b8e0bc} 0.2 & \cellcolor[HTML]{c6e3c6} 0.02 & \cellcolor[HTML]{b8e0bc} 0.2\\
    \hline
    \hline
    \multirowcell{2}[0ex][l]{Average} & F1-Score & \cellcolor[HTML]{b8e0bc} 0.2 (74.7) & \cellcolor[HTML]{a2dbad} 0.5 0.5 (73.4) & \cellcolor[HTML]{ffcbcb} -0.03 (75.2)\\
    & MCC & \cellcolor[HTML]{93d8a2} 0.7 (63.1) & \cellcolor[HTML]{8cd79d} 0.8 (61.1) & \cellcolor[HTML]{a2dbad} 0.5 (64.1) \\
    \hline
    \end{tabular}%
  \label{tab:single_vs_double}%
\end{table}%

\subsection{\mbox{RQ3: Does separating vulnerable statements} and context to provide explicit location of SVs improve assessment performance?}
\label{subsec:rq3_results}

RQ3 evaluated the approach of separating vulnerable statements from their context as two inputs for building SV assessment models. Theoretically, this double-input method tells a model the exact vulnerable and context parts in input code, helping the model capture the information from relevant parts for SV assessment tasks more easily. To separate features, feature vectors are generated for each of the two inputs and then concatenated to form a single vector of twice the size of the vector used in RQ2. RQ3 used the same dataset from RQ2 (i.e., excluding cases without PS context) and the respective model evaluation procedure to objectively compare PS context with the other context types.

\textit{\textbf{Overall, double-input models improved the performance for all types of context compared to single-input ones, but the improvements were not substantial ($\approx$1\%)}}. \tab~\ref{tab:single_vs_double} clearly indicated the improvement trend; i.e., a majority of the cells have green color. We noticed that the rankings of double-input models using different context types still remained the same as in RQ2, i.e., function > PS > surrounding context.
Specifically, double-input models raised the MCC values of single-input models using PS, surrounding, and function context by 1.1\%, 1.4\%, and 0.8\%, respectively. In terms of F1-Score of double-input models, PS and surrounding context had 0.26\% and 0.75\% increase, while function context suffered from a 0.04\% decline. We found the absolute performance differences between double-input and single-input models for the seven tasks were actually small and not statistically significant with negligible effect sizes ($r_{PS~ctx + Vuln} = 0.059$, $r_{Surrounding~ctx + Vuln} = 0.092$, and $r_{Function~ctx + Vuln} = 0.021$). We observed similar changes/patterns of function/surrounding context when using the full dataset in RQ1. The findings suggest that models using function context + Vuln as a single-input in RQ2 still perform competitively. This result strengthens the conclusion in RQ2 that SV assessment models benefit from vulnerable statements along with (in-)directly SV-related lines in functions, yet not necessarily where these lines are located.

\section{Discussion}
\label{sec:discussion}
\subsection{Function-level SV Assessment: Baseline Models and Beyond}
\label{subsec:baseline_and_beyond}

\begin{figure}[t]
    \centering
    \includegraphics[width=\columnwidth,keepaspectratio]{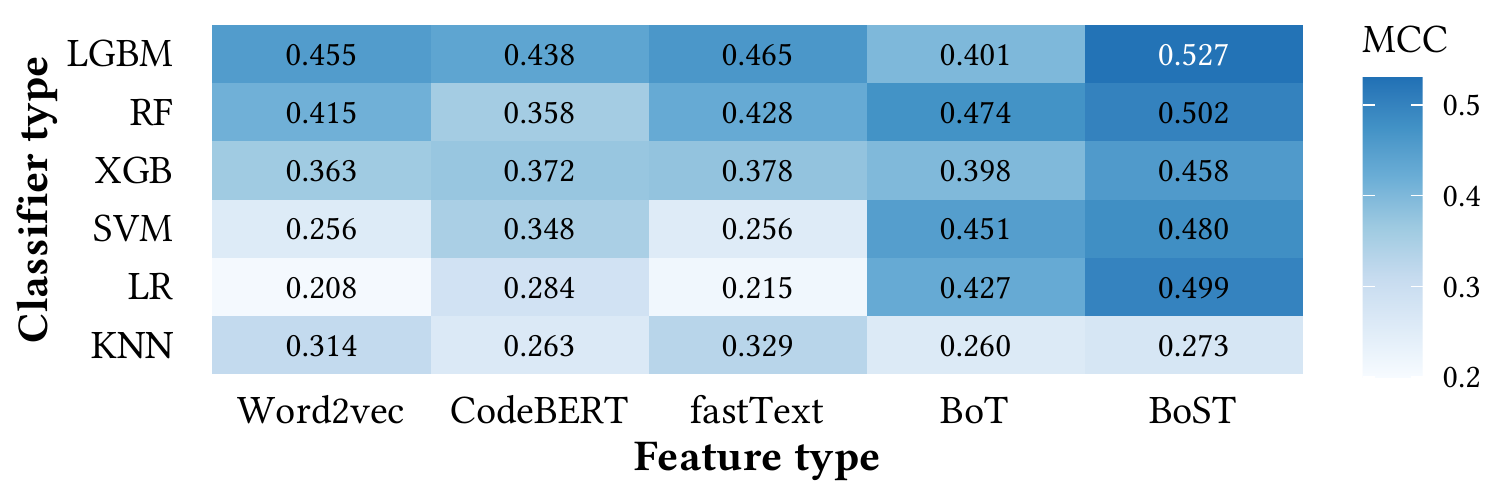}
    \caption{Average performance (MCC) of six classifiers and five features for SV assessment in functions. Notes:
    BoT and BoST are Bag-of-Tokens and Bag-of-Subtokens, respectively.}
    \label{fig:baselines}
\end{figure}

From RQ1-RQ3, we have shown that vulnerable statements and their context are useful for SV assessment tasks. In this section, we discuss the performance of various features and classifiers used to develop SV assessment models on the function level. We also explore the patterns of false positives of the models used in this work. Through these discussions, we aim to provide recommendations on building strong baseline models and inspire future data-driven advances in function-level SV assessment.

\noindent \textbf{Practice of building baselines}. \textit{\textbf{Among the investigated features and classifiers, a combination of LGBM classifier and Bag-of-Subtokens features produced the best overall performance for the seven SV assessment tasks (see \fig~\ref{fig:baselines})}}. In addition, LGBM outperformed the other classifiers, and Bag-of-Subtokens was better than the other features. However, we did not find a single set of hyperparameters that was consistently better than the others, emphasizing the need for hyperparameter tuning for function-level SV assessment tasks, as generally recommended in the literature~\cite{treude2019predicting,tantithamthavorn2018impact}.
Regarding the classifiers, the ensemble ones (LGBM, RF, and XGB) were significantly better than the single counterparts (SVM, LR, and KNN) when averaging across all feature types, aligning with the previous findings for SV assessment using SV reports~\cite{le2019automated,spanos2018multi}. Regarding the features, the ones augmented by sub-tokens (Bag-of-Subtokens, fastText, and CodeBERT) had stronger performance, on average, than the respective feature types using only word-based representation (Bag-of-Tokens and Word2vec). This observation suggests that SV assessment models benefit from sub-tokens, probably because rare code tokens are more likely to be captured by these features. This result is similar to Le et al.~\cite{le2019automated}'s finding for report-level SV assessment models. All of the above comparisons were confirmed statistically significant with $p$-values < 0.01 and non-negligible effect sizes; similar patterns were also obtained for F1-Score.
Surprisingly, the latest feature model, CodeBERT, did not show superior performance in this scenario, likely because the model was originally pre-trained on multiple languages, not only Java (the main language used in this study). Fine-tuning the weights of CodeBERT using SV data in a specific programming language is worthy of exploration for potentially improving the performance of this feature model.
Overall, using the aforementioned baseline features and classifiers, we observed a common pattern in the performance ranking of the seven CVSS metrics across different input types, i.e., Access Vector > Authentication > Severity > Confidentiality -- Integrity -- Availability > Access Complexity. We speculate that the metric-wise class distribution (see \fig~\ref{fig:cvss_distribution}) can be a potential explanation for this ranking. Specifically, Access Vector and Authentication are binary classifications, which have less uncertainty than the other tasks. In addition, Confidentiality, Integrity, and Availability are all impact metrics with roughly similar distributions, resulting in similar performance as well. Access Complexity suffers the most from imbalanced data among the tasks, and thus this task has the worst performance.

\noindent \textbf{False-positive patterns}.
We manually analyzed the incorrect predictions by the optimal models using the best-performing (function) context from RQ2/RQ3. From these cases, we found two key patterns of false positives.
\textit{\textbf{The first pattern concerned SVs affecting implicit code in function calls}}. For example, a feature requiring authentication, i.e., the synchronous mode, was run before user's password was checked in a function (\code{doFilter}), leading to a potential access-control related SV.\footnote{https://bit.ly/32vyggM (CVE-2018-1000134)}
The execution of such mode was done by another function, \code{processSync}, but its implementation along with the affected components inside was not visible to the affected function. Such invisibility hinders a model's ability to fully assess SV impacts. A straightforward solution is to employ inter-procedural analysis~\cite{li2018vuldeepecker}, but scalability is a potential issue as SVs can affect multiple functions and even the ones outside of a project (i.e., functions in third-party libraries).
Future work can leverage Question and Answer websites to retrieve and analyze SV-related information of third-party libraries~\cite{le2021large}.
\textit{\textbf{The second type of false positives involved vulnerable variables with obscure context}}. For instance, a function used a potentially vulnerable variable containing malicious inputs from users, but the affected function alone did not contain sufficient information/context about the origin of the variable.\footnote{https://bit.ly/3plU7QP (CVE-2012-0391)} Without such variable context, a model would struggle to assess the exploitability of an SV; i.e., through which components attackers can penetrate into a system and whether any authentication is required during the penetration. Future work can explore taint analysis~\cite{kim2014survey} to supplement function-level SV assessment models with features about variable origin/flow.

\subsection{Threats to Validity}
\label{subsec:threats}

The first threat concerns the curation of vulnerable functions and statements for building SV assessment models. We considered the recommendations in the literature to remove noise in the data (e.g., abnormally large and cosmetic changes). We also performed manual validation to double-check the validity of our data.

Another threat is about the robustness of our own implementation of the program slicing extraction. To mitigate the threat, we carefully followed the algorithms and descriptions given in the previous work~\cite{dashevskyi2018screening} to extract the intra-procedural backward and forward slices for a particular code line.

Other threats are related to the selection and optimality of baseline models. We assert that it is nearly impossible to consider all types of available features and models due to limited resources. Hence, we focused on the common techniques and their respective hyperparameters previously used for relevant tasks, e.g., report-level SV assessment. We are also the first to tackle function-level SV assessment using data-driven models, and thus our imperfect baselines can still stimulate the development of more advanced and better-performing techniques in the future.

Regarding the reliability of our study, we confirmed the key findings with $p$-values < 0.01 using non-parametric Wilcoxon signed rank tests and non-negligible effect sizes. Regarding the generalizability of our results, we only performed our study in the Java programming language, yet we mitigated this threat by using 200 real-world projects of diverse domains and scales. The data and models were also released to support reuse and extension to new languages/applications~\cite{reproduction_package_msr2022}.

\section{Related Work}
\label{sec:related_work}
\subsection{Code Granularities of SV Detection}
SV detection has long attracted attention from researchers and there have been many proposed data-driven solutions to automate this task~\cite{ghaffarian2017software}. Neuhaus et al.~\cite{neuhaus2007predicting} were among the first to tackle SV detection in code components/files. This seminal work has inspired many follow-up studies on component/file-level SV detection (e.g.,~\cite{chowdhury2011using,shin2013can,tang2015predicting}).
Over time, function-level SV detection tasks have become more popular~\cite{lin2018cross,zhou2019devign,nguyen2019deep,bilgin2020vulnerability,wang2020combining} as functions are usually much smaller than files, significantly reducing inspection effort for developers. For example, the number of code lines in the methods in our dataset was only 35, on average, nearly 10 times smaller than that (301) of files.
Recently, researchers have begun to predict exact vulnerable statements/lines in functions (e.g.,~\cite{li2021vuldeelocator,li2021vulnerability,nguyen2021information, ding2021velvet}). This emerging research is based on an important observation that only a small number of lines in vulnerable functions contain root causes of SVs. Instead of detecting SVs as in these studies, we focus on SV assessment tasks after SVs are detected. Specifically, utilizing the outputs (vulnerable functions/statements) from these SV detection studies, we perform function-level SV assessment to support SV understanding/prioritization before fixing.

\subsection{Data-driven SV Assessment}
SV assessment has been an integral step for addressing SVs. CVSS has been shown to provide one of the most reliable metrics for SV assessment~\cite{johnson2016can}. There has been a large and growing body of research work on automating SV assessment tasks~\cite{le2021survey}, especially predicting the CVSS metrics for ever-increasing SVs. Most of the current studies (e.g.,~\cite{yamamoto2015text,spanos2017assessment,spanos2018multi,le2019automated,elbaz2020fighting,duan2021automated}) have utilized SV descriptions available in bug/SV reports/databases, mostly NVD, to predict the CVSS metrics. However, according to our analysis in section~\ref{subsec:vuln_functions}, NVD reports of SVs are usually released long (up to 1k days) after SVs have been fixed, rendering report-level SV assessment potentially untimely for SV fixing. Unlike the current studies, we propose shifting the SV assessment tasks to the code function level, which can help developers assess functions right after they are found vulnerable and before fixing.
Note that we assess all types of SVs in source code, not only the ones present in dependencies~\mbox{\cite{ponta2018beyond,kritikos2019survey}}.
Overall, our study informs the practice of developing strong baselines for function-level SV assessment tasks by combining vulnerable statements and their context.

\section{Conclusions and Future Work}
\label{sec:conclusions}

We motivated the need for function-level SV assessment to provide essential information for developers before fixing SVs. Through large-scale experiments, we studied the use of data-driven models for automatically assigning the seven CVSS assessment metrics to SVs in functions. We demonstrated that strong baselines for these tasks benefited not only from fine-grained vulnerable statements, but also the context of these statements. Specifically, using vulnerable statements with all the other lines in functions produced the best performance of 0.64 MCC and 0.75 F1-Score. These promising results show that function-level SV assessment tasks deserve more attention and contribution from the community, especially techniques that can strongly capture the relations between vulnerable statements and other code lines/components.

\begin{acks}
This work was supported with supercomputing resources provided by the Phoenix HPC service at the University of Adelaide. We would like to express our gratitude to David Hin for his technical support at the early stage of this work. We also sincerely thank the members from the Centre for Research on Engineering Software Technologies (CREST), Roland Croft, Huaming Chen, Mubin Ul Haque, and Faheem Ullah, as well as the anonymous reviewers for the insightful and constructive comments to improve the paper.
\end{acks}

\balance

\bibliographystyle{ACM-Reference-Format}
\bibliography{reference}

\end{document}